\documentclass[conference]{IEEEtran}
\IEEEoverridecommandlockouts
\usepackage{cite}
\usepackage{amsmath,amssymb,amsfonts}
\usepackage{algorithmic}
\usepackage{graphicx}
\usepackage{textcomp}
\usepackage{url}
\usepackage{multirow}
\usepackage{booktabs}
\usepackage{subcaption}
\usepackage{xcolor}
\def\BibTeX{{\rm B\kern-.05em{\sc i\kern-.025em b}\kern-.08em
    T\kern-.1667em\lower.7ex\hbox{E}\kern-.125emX}}

\usepackage{fancyhdr}
\begin{document}

\title{Towards Understanding Confusion and Affective States Under Communication Failures in Voice-Based Human-Machine Interaction
}

\author{\IEEEauthorblockN{Sujeong Kim}
\IEEEauthorblockA{\textit{SRI International}\\
sujeong.kim@sri.com}
\and
\IEEEauthorblockN{Abhinav Garlapati}
\IEEEauthorblockA{
\textit{Facebook Inc.}\\
abhinavg@fb.com}
\and
\IEEEauthorblockN{Jonah Lubin}
\IEEEauthorblockA{
\textit{University of Chicago}\\
jonahlubin@uchicago.edu}
\and
\IEEEauthorblockN{Amir Tamrakar}
\IEEEauthorblockA{
\textit{SRI International}\\
amir.tamrakar@sri.com}
\and
\IEEEauthorblockN{Ajay Divakaran}
\IEEEauthorblockA{
\textit{SRI International}\\
ajay.divakaran@sri.com}
}

\maketitle
\thispagestyle{fancy}


\begin{abstract}
    We present a series of two studies conducted to understand user's affective states during voice-based human-machine interactions. Emphasis is placed on the cases of communication errors or failures. In particular, we are interested in understanding ``confusion'' in relation with other affective states. The studies consist of two types of tasks: (1) related to communication with a voice-based virtual agent: speaking to the machine and understanding what the machine says, (2) non-communication related, problem-solving tasks where the participants solve puzzles and riddles but are asked to verbally explain the answers to the machine. We collected audio-visual data and self-reports of affective states of the participants. We report results of two studies and analysis of the collected data. The first study was analyzed based on the annotator's observation, and the second study was analyzed based on the self-report.
\end{abstract}


\section{Introduction}\label{sec:introduction}


Voice-base interfaces have become more available with recent advances in natural language processing. Despite the pace of the advances, digital assistants still have shortcomings in understanding the breadth and nuance in natural communication. Commonly reported failures such as general lack of understanding of the users intent, unreliable or inconsistent voice recognition, and inaccurate voice transcription~\cite{Kiseleva:2016:UUS:2854946.2854961} could lead to poor user experience and customer dissatisfaction.

There have been some efforts understanding sources of human-machine communication errors and exploring machine's or human's repair/recovery strategies to handle these errors~\cite{skantze2003exploring_human_error_handling,Bohus05_sorry_i,CwC_HAI_dialog_interface_study,Marge_miscommunication_detection_recovery_2019}. Being able to understand and detect user's affective states in these situations could help develop a more reactive and actively engaging human-machine communication system~\cite{Edlund2008TowardsHS}. There has been extensive work on detecting facial expressions and basic emotions~\cite{Li-FacialExpressionRecognitionSurvey-2020}, such as anger, surprise, happiness, disgust, sadness, and fear, but these categories do not fully describe user states under struggling human-machine communications.

\begin{figure}[ht]
    \centering
    \includegraphics[width=0.9\linewidth]{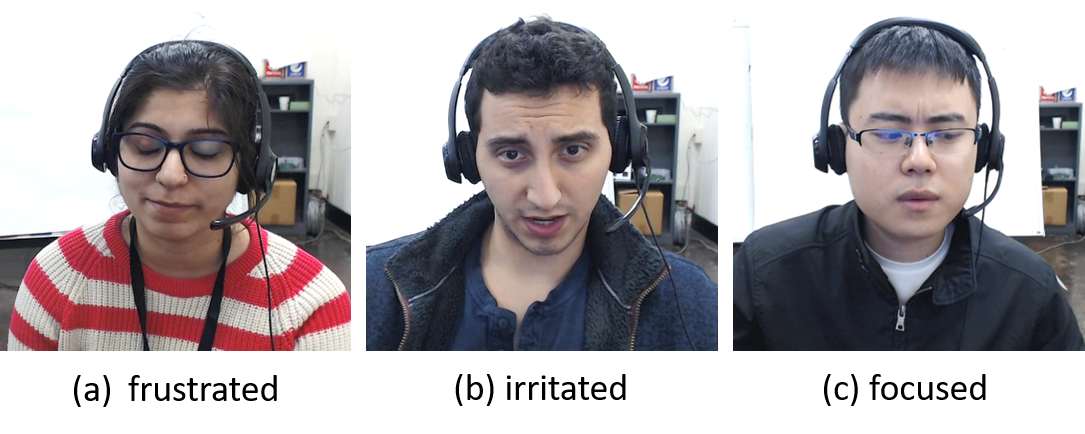}\vspace{-0.5em}
    \caption{\textbf{Example video data} showing different emotions (frustrated, irritated, and focused) exhibited under a highly confused state. Both emotional state and level of confusion are reported by the participants. }\label{fig:confusion-faces}\vspace{-1em}
\end{figure}

In this paper, we present a series of two studies conducted to understand user's affective\footnote{We use the terms \textit{affect} to generally refer to feelings or emotional experience~\cite{barrett2009affect}, interchangeably with the term \textit{emotion}} states during voice-based interactions with machine under existing limitations that could cause communication failures.
Participants are asked to perform two types of tasks: (1) tasks related to human-machine communication, such as speaking to the machine and understanding the machine's speech, (2) problem-solving type tasks, such as puzzles and riddles. We aim to compare user states between these two scenarios and with varying level of difficulties.
We collected multi-modal data (audio-visual recordings) of 35 participants from both studies\footnote{The anonymized data can be shared for research purposes.}.

In the first study, we attempted to categorize commonly exhibited affective states based on the annotations. \textit{Confusion} was the most commonly observed emotion when participants have difficulty completing the task. We found that confusion is exhibited in few different ways in terms of facial/behavior features, which are similar to the features observed in other emotions such as ``frustrated'', ``irritated'', and ``focused''. It led to a follow-up study to understand confusion and other affective states in more depth.
In the second study, we examined user states in three different views, based on the affective state categories revised from the first study, valence and arousal dimensions, and level of confusion, reported by participants after each task. Our results suggest that failures in human-machine communication more often induce confusion than in problem-solving tasks, also accompanying with slightly different emotions (see Figure~\ref{fig:confusion-faces} for example). In the valence and arousal measures, however, we did not find statistically meaningful difference between two types of tasks.


\section{Related Work}\label{sec:relatedWork}

In the area of computer vision and machine learning, most of the existing work on emotion/affect detection has been focused on basic emotions or dimensional models~\cite{Li-FacialExpressionRecognitionSurvey-2020}. Due to its wide application, it has been an active area of research in other fields, too, including robotics~\cite{robotAssistantsSurvey}, human-computer interaction~\cite{McintyreCompositeSensingOfAffect} and learning~\cite{survey-emotion-recognition-e-learning-2019-imani,affect-detection-interdisciplinary-review}, and focuses more on detecting emotions in by uni-/multi-modal data in natural settings.

There have been studies on user's affective/emotional states outside the basic emotion categories. Some studies suggest agents responding to user's feelings reduce user's frustration~\cite{Klein2002ThisComputerResponds,Hone2006EmpathicAgents}. Hoque et al.~\cite{Hoque_2015_frustration} propose a model to distinguish frustration and delight, Ishimaru et al.~\cite{ishimaru2021confidenceaware} propose a learning assistant that gives feedback based on self-confidence detection, Halfon et al.~\cite{psychotherapy-2020-halfon} present a tool to analyze anger, anxiety, pleasure, and sadness in psychotherapy.
There have been some studies about understanding learner's states including confusion as a cognitive-affective state~\cite{dmello2014confusion} and integration with an affect-sensitive tutor~\cite{AffectSensitiveTutor}. Some studies focus on using these algorithms to study the effect of confusion on learning performance, for example, how confusion and frustration increase the performance of learning~\cite{Liu2013SequencesOF}.

\section{Method}\label{section:methods}

We designed a web-based program where participants can interact with the computer using a voice interface. We used the WebKit Speech API \cite{WebkitSpeechAPI} to transcribe the participants' utterances to pass to the system and to synthesize machine speech. For data collection, we set up a typical computer desk: computer with a monitor, a keyboard, a mouse and a headset. We installed a web camera on top of the monitor so it can capture the face and the upper body of a participant. Recorded data contain the task screen and the web-camera input. Additionally, we collected a survey for the first study and self-report for the second study.

The study starts with a short tutorial of the voice interface and presents four sessions, two containing tasks related to communication with the voice agent, the other two containing problem-solving type tasks:

\paragraph{Speak to the machine (communication)}
This session is related to the system's natural language understanding ability, more specifically, correctly taking in user's utterances. Participants are asked to speak a pre-selected sentence to the machine. For difficult tasks, sentences were chosen such that a typical automatic speech recognition (ASR) system would have trouble transcribing.

\paragraph{Repeat after the machine (communication)}
This session is related to system's speech synthesis, more specifically, naturally and clearly delivering speech to the users. Participants are asked to repeat the sentences spoken by the machine. All the sentences were in English, but machine speech is generated with various foreign accents.

\paragraph{Puzzles (problem solving)}
Participants are asked to solve puzzles with varying difficulty. They must speak out their solution to the machine.

\paragraph{Riddles (problem solving)}
The voice agent shows an image and tells a riddle related to the image. Participants must speak out their solution to the machine.

The tasks cover range difficulty. Participants were given three chances until they complete the task correctly. There was no time limit for any tasks. It took about 40 minutes to finish the study including training and survey/self-report.








\vspace{-0.1em}
\section{Study1: Observed User States}

We collected data from 25 participants, $9$ females and $16$ males, age between $20$ to $69$. $36\%$ of them reported that they almost never use any voice-assistants like Siri, Alexa, or Google Assistant, $24\%$ use almost everyday, $12\%$ few times a year, $8\%$ few times a month, and $20\%$ few times a week.


Participants are given a brief survey after completing all tasks. For the question asking if they were confused at any circumstances during the study, 76\% answered yes. Out of them, 75\% indicated that it was during the communication tasks, 17\% during the problem-solving tasks.

The data was annotated by three annotators. Annotators were asked to watch the recordings, segment and label any emotional expressions exhibited. We were particularly interested in finding user state categories that might not be describable with six basic emotion categories, so we asked the annotators to use their judgement and pick a label that best describes each segment. When we merged the annotations, we picked the majority annotations for overlapping segments. When no such majority annotation existed, the more experienced annotator's annotation was chosen. We extracted 2004 video segments in total, with lengths from 1 second to several minutes. Labels suggested by the annotators were: neutral (218), happy (333), surprise (302), apprehensive/anxious (60), confused (730), frustrated (64), irritated (144), disappointed (101), thoughtful (38), realization (1), dissatisfied (13).

Some facial/behavioral features that are often observed under challenging tasks were: furrowed eyebrows, raising one or both eyebrows, closing or blinking eyes, pressing lips, tilting head, changing eye gaze from computer screen to outside the screen (left, right, down, or up), getting closer to the screen or speaker to listen closer, leaning forwards or backwards while staring at the screen, and resting chin on hand. About 72\% of extracted segments contain more than two features exhibited, while the rest contain only one of the features.

\begin{table*}[ht]
\resizebox{\linewidth}{!}{%
    \begin{tabular}{@{}lllc@{}} \toprule
        Session                         & Top rated challenge(s)                                                      & Top rated main emotion(s)       & \multicolumn{1}{l}{Avg. confusion} \\ \midrule
        \multirow{2}{*}{SpeakToMachine} & NONE or MINOR: The task was easy (50\%)                                     & Indifferent                     & \multirow{2}{*}{1.66}              \\
                                        & ASR SYSTEM FAILURE: The computer did not work as I expected (43\%)          & Unexpected, negative surprise   &                                    \\ 
        RepeatAfterMachine              & MACHINE SPEECH: I had difficulty understanding what the computer was saying & Irritated, annoyed              & 2.69                               \\ 
        Riddles                         & PROBLEM SOLVING: The question was difficult to solve/answer                 & Unsure, uncertain, apprehensive & 2.33                               \\ 
        Puzzles                         & PROBLEM SOLVING: The question was difficult to solve/answer                 & Unsure, uncertain, apprehensive & 2.96                               \\ \bottomrule
    \end{tabular}%
        }
        \caption{Top rated challenge(s), emotion(s), and average level of confusion for each session (different types of tasks). Note that ASR system failure and machine speech failure are categorized as communication failure for analysis, but reported separately in this table.}\label{tab:per_session_challenge_main_emotion}\vspace{-1em}

\end{table*}

One of the main challenges in annotation was due to individual difference in their neutral face expression, which serves as the baseline for determining exhibition of other emotions. Different facial appearance, especially related to shape of the lips and eyebrows, made it challenging to distinguish between neutral state and exhibition of certain emotions. Annotating data from a wide age range of participants was also challenging for similar reasons~\cite{aging_malatesta1987affect,aging_and_FER}.


\textbf{Revised user state categories}: We reviewed the extracted clips without context, i.e., only a segment of a video recording without task information or audio, thus solely based on the aforementioned facial/behavioral features
. Some clips labeled confused were difficult to distinguish from other categories like irritated, frustrated, apprehensive/anxious, or focused, because of similarity of the facial/behavioral features. Because of this ambiguity, we decided to separate out confusion and further analyze it in relation with other affective states in the follow-up study.
We also revised the user state categories to cover subtle differences in affective states under challenging tasks. The revised categories are: indifferent, happy/amused, surprise/realization (e.g., ``oh, I got it!, wow!''), negative surprise/unexpected (e.g., ``what’s going on?''), focused/thoughtful, uncertain/apprehensive/anxious, irritated, and frustrated.

\begin{figure*}[ht]
    \centering
    \includegraphics[width=0.9\linewidth]{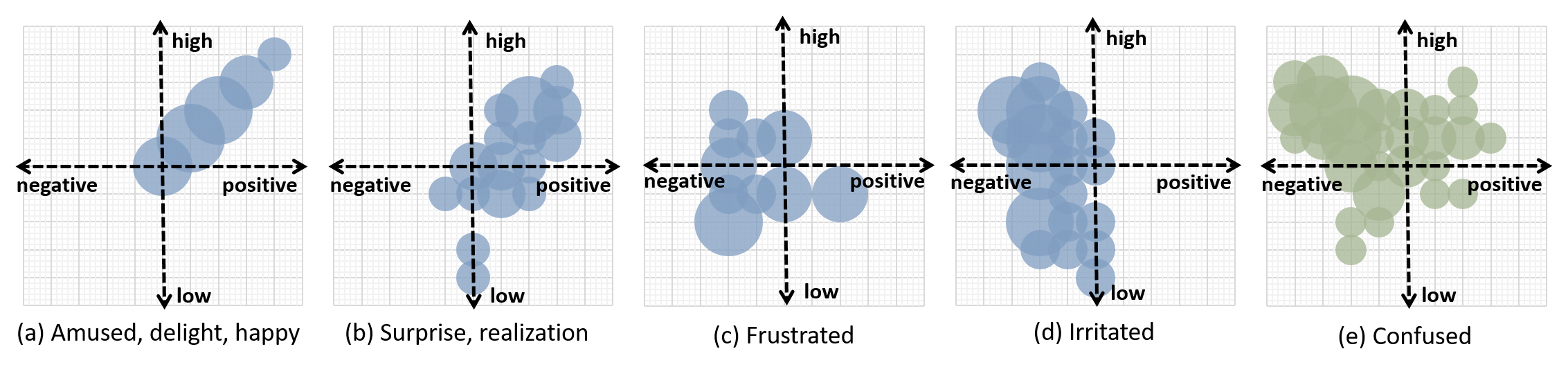}
    \vspace{-0.5em}
    \caption{Valence (negative/positive) and arousal (low/high) measures for selected emotional states (a)-(d) and high level of confusion (e). Size of the circle is proportional to the data count. (a) and (b) are distributed mostly in the positive valence area. (c),(d),(e) are mostly concentrated in the negative valence area, with (e) having some data appear in positive valence area.
    }\label{fig:emotion-confusion-affect}
\end{figure*}

\begin{table*}[]
    \centering
    \begin{tabular}{cccccccccc}
        \toprule
        Confusion & Unsure & Unexpected & Surprise & Amused & Focused & Frustrated & Indifferent & Irritated & Total \\ \midrule
        1                          & 3.74\% (4)              & 4.68\% (5)                  & 8.42\% (9)                & 24.3\% (26)             & 9.35\% (10)              & 3.74\% (4)                  & 38.32\% (41)                 & 7.48\% (8)                 & 100\% (107)            \\ 
        2                          & 17.08\% (7)             & 14.64\% (6)                 & 19.52\% (8)               & 12.2\% (5)              & 4.88\% (2)               & 7.32\% (3)                  & 9.76\% (4)                   & 14.64\% (6)                & 100\% (41)             \\ 
        3                          & 7.9\% (3)               & 26.32\% (10)                & 15.79\% (6)               & 7.9\% (3)               & 7.9\% (3)                & 7.9\% (3)                   & 13.16\% (5)                  & 13.16\% (5)                & 100\% (38)             \\ 
        4                          & 22.23\% (8)             & 8.34\% (3)                  & 2.78\% (1)                & 2.78\% (1)              & 11.12\% (4)              & 25\% (9)                    & 5.56\% (2)                   & 22.23\% (8)                & 100\% (36)             \\ 
        5                          & 16.67\% (3)             & 16.67\% (3)                 & 5.56\% (1)                & 0\% (0)                 & 22.23\% (4)              & 27.78\% (5)                 & 0\% (0)                      & 11.12\% (2)                & 100\% (18)             \\ 
        Total                      & 10.42\% (25)            & 11.25\% (27)                & 10.42\% (25)              & 14.59\% (35)            & 9.59\% (23)              & 10\% (24)                   & 21.67\% (52)                 & 12.09\% (29)               & 100\% (240)            \\ \bottomrule
    \end{tabular}\caption{Percentage and count of each emotional states (used short-terms for the table) reported for different levels of confusion. Confusion levels range from 1 (not confused at all) to 5 (extremely confused). 
    }\vspace{-0.7em}
    \label{tab:emotion-for-all-confusion}
\end{table*}

\section{Study 2: Self-report Analysis}\label{sec:study2}

In the second study, we addressed research questions that arose from the first study by evaluating the user state categories. This time, we collected a self-report from participants after each task about their emotional states, affect state in terms of valence and arousal, level of confusion, and the main challenge of the task.
The questionnaire included the following questions and options to choose from.
\paragraph{Main challenge} ``none or minor'', ``ASR system failure'', ``machine speech'', ``problem solving'', ``memory'', ``other'', with short descriptions for each option.
\paragraph{Most significant feelings/emotions during the task} ``indifferent'', ``happy/amused'', ``surprise/realization'', ``unexpected/negative surprise'', ``unsure/uncertain/apprehensive'', ``irritated/annoyed'', ``frustrated'', ``focused/thoughtful'', ``other''. A follow-up question asks to select all other significant feelings/emotions experienced. Same options plus ``no other feelings'' are given.
\paragraph{How pleasant the experience was during the task (valence), and how strong it was (arousal)} Self-Assessment Manikin (SAM)~\cite{SAM-Lang1985TheCP} in scale of 1 to 9, each.
\paragraph{Level of confusion} in scale of 1 (not confused at all) to 5 (extremely confused).

We collected total 241 sets of self-report from 10 participants, 5 females and 5 males. 9 of them reported their age range 20-29 (45\%), 30-39 (9\%), 40-49 (36\%), 50-59 (9\%). 33\% of them answered that they have almost never used voice assistants, 33\% answered they use almost everyday, 11\% few times a week, 11\% few times a month, 11\% few times a year.

\subsection{Emotional states}

Table~\ref{tab:per_session_challenge_main_emotion} shows the top rated challenge(s), emotion(s), and average level of confusion for each session obtained from self-reports. The ``Speak to Machine'' session was considered relatively easy because of the high accuracy of the ASR engine. The second ranked challenge for this session was ASR system failure, with associated main emotion(s) unexpected/negative surprise. For other sessions, only the top-ranked items are shown in the table. We can see different main emotions are reported for different type of challenges. Interestingly, challenges in human-machine communication tasks induce more negative emotions than those reported in problem-solving tasks.

In Table~\ref{tab:emotion-for-all-confusion}, the last row shows the distribution of main emotions reported from all participants. All categories are fairly equally distributed except indifferent at about twice the others. We compare these emotional states in valence-arousal dimensions. We visualize valence-arousal scores for selected emotions in Figure~\ref{fig:emotion-confusion-affect} (a)-(d). We can see that amused/happy and surprise/realization are trending positively, whereas frustrated and irritated are more distributed in the negative space along the valence axis. Not included in the figure, other emotions such as unexpected/negative surprise, unsure/uncertain/apprehensive, focused/thoughtful, do not show clear trend in distribution either in negative or positive valence axis. Based only on this result, these states might not be suitable as distinctive affective states, but we believe they could still provide important cues for human-machine interaction. Further analysis in facial/behavioral/audio features from the recordings could help better understand the emotional states.

\vspace{-0.5em}
\subsection{Level of confusion}

The average level of confusion for incomplete tasks, i.e., participants failed to correctly perform the task after three trials, was 2.93, which was higher than average level of confusion, 1.67, for complete tasks ( $p\ll0.5$, two-sample t-test assuming unequal variance).

In the first study, we observed ambiguity in the presentation of confusion, sharing similar facial/behavioral features from other user state categories. Table~\ref{tab:emotion-for-all-confusion} shows distribution of the main emotions for different level of confusion in range 1 (not confused at all) to 5(extremely confused). In the following analysis, we refer to levels 4 and 5 as ``high confusion``. Under high confusion, emotions such as frustration (25\%), unsure, uncertain, apprehensive 20\%, irritated 18\%, focused, thoughtful 15\%, unexpected, and negative surprise 11\% were reported more often than others.

Figure~\ref{fig:confusion-faces} shows screenshot of three participants under high confusion state. The main emotions they reported were: frustrated, irritated, and focused, respectively. We can see the difference in the exhibited emotions from their facial/behavioral expressions, such as eye gaze, and shape of the lips and the eyebrows.
This confirms the observation in the first study that confusion is exhibited in multiple different forms.

In the valence-arousal dimension (see Fig.~\ref{fig:emotion-confusion-affect} (e)), confusion is mostly distributed in the negative valence and high arousal space, but we observe some data points in other quadrants. The positive ones include main emotions as surprise/realization, indifferent, unsure/uncertain and unexpected, mostly with amused/happy reported as secondary emotion. This suggests that confusion can sometimes be positive experience based on the outcome, but we leave further analysis as future work.

Next, we compare high confusion between human-machine communication tasks and problem-solving tasks. Our results show that human-machine communication tasks generally induced confusion more often than for problem-solving tasks (60\% vs. 40\%), also in failure cases (55\% vs. 45\%). Under highly confused states, the two most frequently reported main emotions were frustrated (28\%) and irritated (21\%) for human-machine communication tasks, and unsure/uncertain/apprehensive (32\%) and frustrated (23\%) for problem-solving tasks.

In the valence-arousal measures, we did not find statistically meaningful difference ($p>0.5$, two-sample t-test assuming unequal variance) between the two types of tasks. This indicates that based on the data, there is not enough evidence to tell if confusion is exhibited differently between the two types of tasks in terms of valence-arousal measures. A follow-up larger scale study might help confirming this result.


Based on the results, we think that confusion is not an independent emotional state with unique facial/behavioral features or valence/arousal measures. Rather, it can be considered as a higher level user state that we can infer from exhibition of other emotions over time and based on the context.
Our results suggest that participants were more often confused and felt negative towards failures in human-machine communication tasks than those in problem-solving tasks. Thus, preventing the communication failures or having a way to recover from these failures might help greatly improving user experience.

\section{Conclusion and Future Work}
We presented results of two studies on user states in voice-based human-machine interaction. In the first study, we analyzed the data based on annotations and suggest user state categories. In the second study, we analyze user states based on self-report, in terms of discrete emotional categories identified in the first study, valence/arousal measures, and level of confusion.
Confusion is often induced when human-machine communication fails, more specifically due to failure of the machine to understand human speech or failure to generate easily understandable machine speech. We found that confusion in these scenarios is a complex state where different emotion can be exhibited. We also found some difference in emotions towards failures in human-machine communication tasks compared to problem-solving tasks, but no statistically meaningful difference between confusion in two different types of tasks in valence/arousal measures.

As future work, we would like to look at audio-visual features and compare with the self-report analysis. We are planning on training a machine learning model to detect the user states and perform a follow-up study to learn if early intervention in confusion state would improve user experience or help task completion. We are also interested in incorporating individual differences and context information~\cite{HoqueKP09_Disagree_Facial_Affect}.

\paragraph*{Acknowledgment}

The authors would like to thank Edgar Kalns, Elizabeth Shriberg and Andreas Kathol for their support and feedback throughout the project.

\clearpage
\bibliographystyle{IEEEtran}
\bibliography{references}

\end{document}